\documentclass[pre,aps,twocolumn,groupaddress,longbibliography]{revtex4-1}
\usepackage{latexsym}
\usepackage{lineno}
\usepackage{hyperref}
\usepackage{url}
\usepackage{graphicx}
\usepackage{verbatim}
\usepackage{multirow}
\usepackage{amsmath}
\usepackage{SIunits}
\newcommand{\beq}{\begin{eqnarray}}
\newcommand{\eeq}{\end{eqnarray}}
\usepackage{mathrsfs}
\usepackage{float}
\usepackage[usenames, dvipsnames]{color}
\usepackage{mathtools}
\usepackage{slashed}
\usepackage{graphicx}   
\usepackage{epstopdf}
\usepackage{subfigure}  
\usepackage{hyperref}   
\usepackage{bbold}
\usepackage{wasysym}
\usepackage{feynmp}
\usepackage[symbol]{footmisc}

\usepackage[latin1]{inputenc}
\usepackage{tikz}
\usetikzlibrary{decorations.pathmorphing}
\usetikzlibrary{shapes.misc}
\tikzset{cross/.style={cross out, draw=black, minimum size=8*(#1-\pgflinewidth), inner sep=0pt, outer sep=0pt},
cross/.default={1pt}}
\usetikzlibrary{patterns,math}
\begin{document}

\title{Analytical theory of enhanced Bose-Einstein condensation in thin films}

\author{\textbf{Riccardo Travaglino}$^{1}$}%
\author{\textbf{Alessio Zaccone}$^{1,2}$}%
 \email{alessio.zaccone@unimi.it}
 
 \vspace{1cm}
 
\affiliation{$^{1}$Department of Physics ``A. Pontremoli'', University of Milan, via Celoria 16,
20133 Milan, Italy.}
\affiliation{$^{2}$Cavendish Laboratory, University of Cambridge, JJ Thomson
Avenue, CB30HE Cambridge, U.K.}

\begin{abstract}
We present an analytically solvable theory of Bose-Einstein condensation in thin film geometries. Analytical closed-form expressions for the critical temperature are obtained in both the low-to-moderate confinement regime (where the film thickness $L$ is in the order of microns) as well as in the strong confinement regime where the thickness is in the order of few nanometers or lower. The possibility of high-temperature BEC is predicted in the strong confinement limit, with a square-root divergence of the critical temperature $T_{c} \sim L^{-1/2}$. For cold Bose gases, this implies an enhancement up to two orders of magnitude in $T_c$ for films on the nanometer scale. Analytical predictions are also obtained for the heat capacity and the condensate fraction. A new law for the heat capacity of the condensate, i.e. $C \sim T^{2}$, is predicted for nano-scale films, which implies a different $\lambda$-point behavior with respect to bulk systems, while the condensate fraction is predicted to follow a $[1- (T/T_c)^{2}]$ law.
\end{abstract}
\maketitle

\section{Introduction} 
Bose-Einstein condensation was predicted early on based on statistical mechanics, although its experimental discovery had to wait until modern technology allowed experimentalists to reach low temperatures in the order of 100 nanoKelvins \cite{first_BEC,stringari}. 

Much research has been directed since then towards finding systems where BEC can be observed at higher temperatures.
Solid-state quasi-particles and photons in microcavities can form BEC condensates at much higher temperatures due to their very light or vanishing masses \cite{magnonsPRL}. The highest BEC critical temperatures (at room temperature) have been reached with photons in dye-filled microcavities \cite{Schmitt2010}, and subsequently also with excitonic systems \cite{Wang2019}.

Ways of increasing the critical temperature $T_c$ with bosonic atoms are rather limited, although early research by the Russian school in the area of superconductivity and superfluidity highlighted the effect of spatial size confinement as an effective way of reaching higher $T_c$s for superfluids and superconductors, see e.g. \cite{Ginzburg1968}.
In particular, the thin film geometry is very promising because it allows one to effectively increase the $T_c$ by restricting the available states at low energy \cite{BEC_GrossmannHolthaus}, while at the same time allowing one to keep the system macroscopically large (which would be impossible with confinement increasing in more than one spatial directions). 

Much research over the past decades has also been directed to studying the properties of atomic monolayers and multi-layers of helium adsorbed on graphite or similar carbon-based surfaces \cite{Saunders}, in terms of structure and dynamics, as well as onset temperature of superfluidity \cite{Reppy,Williams_onset}, and similar studies are available for two-dimensional (2D) optical Bose systems \cite{Schmitt2010}. In this two-dimensional (2D) limit \cite{Blatter}, the onset of superfluidity is a very complicated problem due to various factors, including two-phase coexistence, gas-liquid transition, possible supersolidity of the second layer \cite{PRL2021}, various 2D transitions similar or related to the Berezinskii-Kosterlitz-Thouless (BKT) transition. Approaches based on BKT theory predict the onset temperature of superfluidity to grow with film thickness in this regime \cite{Bishop1980}, which is well supported by experimental data \cite{Williams_onset}, although re-entrant behaviours and non-monotonicity are also well documented \cite{Saunders,PRL2021}. It should be noted that, according to the Mermin-Wagner theory, there is no BEC for free particles in 2D systems.
There is, however, in the presence of a confining harmonic potential trap \cite{Bagnato,Phillips2D,Littlewood_polaritons}.
In finite-size homogeneous 2D systems, however, BEC is re-established once the phase correlations extend over the entire system, see e.g. \cite{Hadzibabic2011} and recent experimental work, \cite{Schmitt2021}.
In particular, \cite{Schmitt2021} has for the first time studied the relationship between critical temperature (or particle number) and system size in a Bose gas. Further work that is closely related, focused on exploring the dimensional crossover from 2D to 1D in the context of specific heat, has been reported very recently \cite{Stein2022}. 

Furthermore, although the BKT phenomenology is active in 2D and may have some interplay with BEC, the two transitions are well distinct phenomena because BKT requires interactions whereas BEC only quantum statistics.

While the 2D limit has been extensively studied, the remaining very broad range of film thickness from sub-millimeter down to the nanometer scale has remained surprisingly unexplored. A big open question here is at which length scale along the confined direction the BEC starts to appear. While the Mermin-Wagner theory states that no BEC would occur at exactly $d=2$ for extended free-particle systems, calculations coming from the 3D limit of finite films like the one reported here or in \cite{BEC_confinement} for the cubic-box geometry, suggest a divergence of $T_c$ as the thickness goes to zero, asymptotically. The emerging picture, to be tested in future work, is that a maximum in $T_c$ could occur for some very small but finite length scale along the confining direction, before $T_c$ drops to zero according to the Mermin-Wagner theory, as the exactly 2D limit is reached.

Hence, in this paper, we focus on this largely unexplored problem and we develop the first systematic and fully analytical theory of confinement effects on BEC in thin films covering a broad range of thickness from sub-millimeter down to the nanometer scale. The theory is based on rigorously accounting for the deformation of the available momentum space induced by the confinement along one spatial direction. Analytical predictions in closed-form are presented for the $T_c$ as well as for the heat capacity, in a broad range from sub-millimeter confinement to nanometer scales using two different perturbative expansions. Experimentally testable predictions are discussed as well as their implications for high temperature BEC and superfluidity.

\section{Theoretical framework} 
\subsection{Confinement geometry in momentum space}
Confinement of a quantum system causes its fundamental properties to change, because of the redistribution of the accessible states in momentum space. Numerical models for the full wave propagation problem can be formulated, by considering a variety of different boundary conditions (BCs), such as periodic or Dirichlet BCs (the so called ``hard walls'' BCs). However, numerical solutions often overshadow the physical mechanisms, so that it is desirable to have analytically tractable theories. 
Furthermore, BCs used in numerical treatments often entail arbitrary assumptions about the behaviour at the physical borders of the system, and are ultimately irrelevant for comparison with experimental data \cite{omega3,BEC_confinement}.

Also, our theory applies to systems with spatial dimensions $d>2$, i.e. for thin films where motions and vibrations are still possible also in the confined direction, and hence the limit $L \rightarrow 0$ has to be taken only as an asymptotic limit.

\begin{figure}[ht]
    \centering
    \includegraphics[width=.45\textwidth]{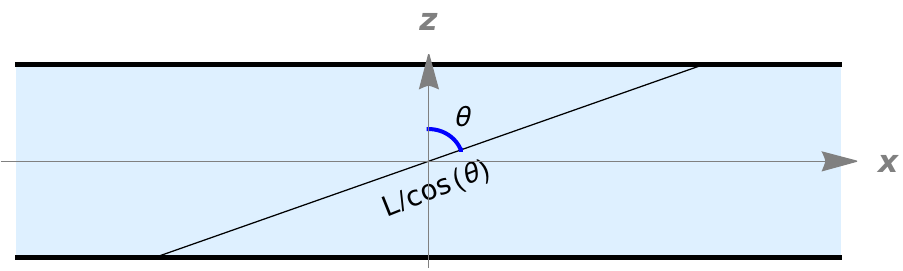}
    \caption{2D section of a thin film of thickness $L$, confined along $z$ and infinite along the $y$ and $x$ directions. A Bose gas atom (quantum plane wave) is assumed to have a maximum wavelength equal to the length of the medium in the direction of motion, which can be expressed as a function of the angle $\theta$, owing to the cylindrical symmetry, as $\lambda_{max} = L/\cos{\theta}$. This leads to a cutoff in the accessible values of wavevector $k$. }
    \label{fig:my_label}
\end{figure}

In this work, following the ideas from references \cite{conf1, conf}, we consider a system (thin film) confined in the $z$-direction, as shown in Fig. \ref{fig:my_label}, and unconfined in the $x$ and $y$ directions. The following discussion is hence directly relevant to the study of the physics of thin films of cold atoms and superfluids. In order to perform calculations, atoms are treated as quantum plane waves as in the standard Bose gas model, with energy $\epsilon=\frac{\hbar^2k^2}{2m}$ \cite{pahria1977,huang}. The cylindrical symmetry of the system allows one to characterize the states in momentum space of a particle using only the angle $\theta$.
In our analytical treatment we do not need to impose any arbitrary BCs, which anyway are irrelevant for final results \cite{BEC_confinement,omega3}. 

The effect of confinement is taken into account by setting a cut-off in the accessible low-energy states, by recognizing that the free quantum particles (or, equivalently, quantum waves) moving in a direction defined by the angle $\theta$ can have a maximum possible wavelength given by~\cite{conf}:
\begin{equation}
    \lambda_{max} = \frac{L}{\cos\theta}.
\end{equation}
This condition implies that the wavelength of a quantum particle cannot exceed the extension of the sample along a particular direction. This is clear from Fig.~\ref{fig:my_label}. 
Since the wavelength of a particle is related to its wavenumber by the relation $\lambda=\frac{2\pi}{k}$, this condition is equivalent to a cutoff condition on the minimum possible wavenumber that can be associated with the free particle:
\begin{equation}
 k_{min} = \frac{2\pi \cos{\theta}}{L}.
 \label{eq:kmin}
\end{equation}
Upon considering plane wave states that propagate in the real-space material depicted in Fig.\ref{fig:my_label}, it is possible to analytically calculate the geometry of the corresponding volume in momentum space. This was done in Ref.\cite{conf} for phonons/elastic waves and the result for phonons is summarized in Fig. \ref{fig:spheres}.
The condition shown above can be used to select the correct lower limit of integration to obtain the available volume in momentum space corresponding to a given volume in real space. 

\subsection{The example of phonons}
\begin{figure}[ht]
    \centering
    \includegraphics[width=.4\textwidth]{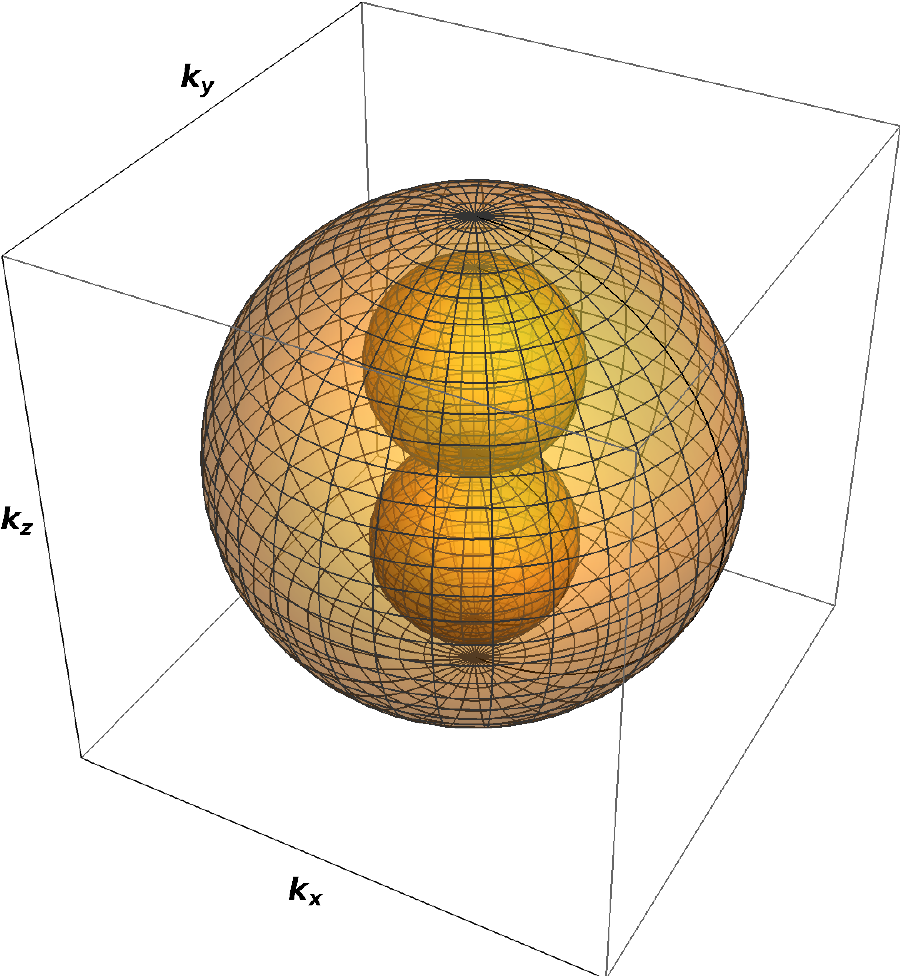}
    \caption{The allowed momentum space for phonon plane waves propagating in a confined sample with the geometry sketched in Fig. \ref{fig:my_label}, as calculated analytically in Ref.\cite{conf} for phonons. The two inner spheres represent the set of forbidden states in $k$-space, while the outer sphere is the Debye sphere for the bulk material. The volume of available states in $k$-space is represented by the volume of the outer Debye sphere minus the volumes of the two smaller spheres which represent states that are not available due to confinement.}
    \label{fig:spheres}
\end{figure}

In particular, as demonstrated exactly with analytical derivations in Ref.\cite{conf} for acoustic phonons, the condition Eq.\eqref{eq:kmin} for the sample geometry of Fig.\ref{fig:my_label}, identifies two spheres of ``forbidden states'' in momentum space, both of radius $\frac{\pi}{L}$ centered in $(0,0,\pm \frac{\pi}{L})$, as shown in Fig.~\ref{fig:spheres}. In the figure, the outer (Debye) sphere, of radius $k_{D}$, which represents all allowed states for plane waves in a bulk unconfined material is shown together with the two ``hollow'' spheres representing states that are forbidden due to the confinement. Therefore, when converting sums over wave vectors to integrals over the available momentum space, the integrals must not be carried over the whole Debye sphere, as standard for phonons in unconfined materials, but rather on the manifold given by the Debye sphere minus the two spheres of forbidden states.

\subsection{The case of the Bose gas}
In the following we are interested in describing the effect of confinement on the geometry of available momentum space for Bose gas particles, within the typical assumptions of BEC in cold gases \cite{huang}. 

What changes in the case of the Bose gas, with respect to the example of phonons, is that there is no maximum energy of allowed states. This is of course very different from the case of phonons, where the Debye frequency $\omega_{D}$ is strictly an insurmountable limit. 
Hence, in the case of phonons, the two spheres of forbidden states can grow, upon increasing $L$, only up to the point where they are just touching the Debye wavevector $k_D$. They cannot grow any further than that.

In the Bose gas, the occupation of states above the ground state is an effect entirely due to finite temperature, which depends on the factor $\frac{1}{\exp{\beta (\epsilon-\mu)}-1}$, where $\mu$ is the chemical potential, and on the density of states (DOS), $g(\epsilon)$. There is no cap imposed by an ultraviolet cut-off such as $\omega_D$ for phonons. 
Hence, in this case, upon increasing the confinement (i.e. upon decreasing $L$) the main consequence is that there will be a region in momentum space where the DOS is not given by the usual DOS of the Bose gas, which is $g(\epsilon)= \frac{V(2m)^{3/2}}{(2\pi)^2\hbar^3} \sqrt{\epsilon}$, but has a different form that will be derived in the next sections.
In turn, due to this different form of the DOS, the integral over the Bose-Einstein (BE) distribution will give a different result in that region of momentum space. In particular, as we shall see later on, since many states at low energy become forbidden because of the confinement, occupation of the ground state now occurs at a higher temperature. Otherwise, the only alternative would be that states with higher energy were to be occupied, which however has a much lower probability due to the Boltzmann factor. 

In the next sections we will explore the above considerations quantitatively, starting from the modified DOS, in order to arrive at an analytical theory of BEC which takes into account the redistribution of states in momentum space induced by the confinement.

\section{Density of states}
The number of allowed low energy states, in general, will be different in the presence of confinement. Hence, the density of states (DOS) will have a different structure than the traditional DOS for the Bose gas. The DOS as a function of energy can be written as:
\begin{equation} 
 g(\epsilon) = \frac{d}{d\epsilon} N(\epsilon'<\epsilon),
\end{equation}
where $N(\epsilon'<\epsilon)$ is the number of states with energy smaller than $\epsilon$.

\begin{figure}[ht]
    \centering
    \includegraphics[width=.5\textwidth]{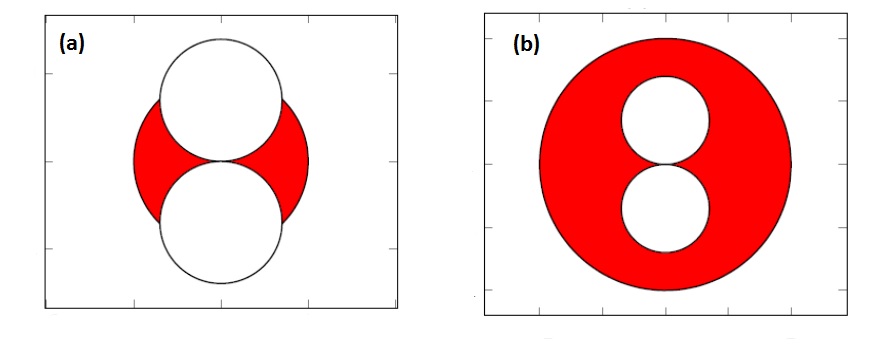}
    \caption{2D section, in the $k_{x}$-$k_{z}$ plane, of the volume of allowed states in $k$-space (red area) for a Bose gas. The DOS must be calculated considering only the highlighted red zone of available states. In (a), $k<\frac{2\pi}{L}$, and the derivative of the number of states used to evaluate the DOS must take the two (white) spheres of forbidden states into account. In (b), $k>\frac{2\pi}{L}$, and therefore the derivative used to evaluate the DOS is unaffected by the forbidden states. One should note that, for $L=const$, (a) would also correspond to a low temperature (internal energy < energy associated with the confinement) situation, while (b) would correspond to the opposite case of high temperatures. }
    \label{fig:Fig3}
\end{figure}

Fixing some energy $\epsilon$, it is thus necessary to count the number of states with lower energy. In order to do so, it is convenient to work in $\mathbf{k}$-space where the DOS of a Bose gas is simply $\frac{V}{(2\pi)^3}$ ($V$ is the volume of the sample in real space).

The number of states with momentum lower than $k$ is given by:
\begin{equation}
N(k'<k) = \frac{V}{(2\pi)^3} Vol_k    
\label{eq:N}
\end{equation}
where $Vol_k$ is the available volume in $\mathbf{k}$-space. 

There are two different possibilities depending on whether $k<\frac{2\pi}{L}$ or $k>\frac{2\pi}{L}$. The latter case is trivial: one has
\begin{equation}
    Vol_k = \frac{4}{3} \pi k^3 - 2 \frac{4}{3} \pi \left(\frac{\pi}{L}\right)^3.
\end{equation}
The volume of two forbidden spheres of radius $\frac{\pi}{L}$ is then subtracted from the volume of the $k$-sphere (valid for an unconfined system), which returns the volume of available states in $k$-space. Because this correction does not depend on $k$, it does not affect the derivative and hence does not affect the DOS. The $g(\epsilon)$ is therefore the usual DOS for the Bose gas:
\begin{equation}
g(\epsilon)= \frac{V (2m)^{3/2}}{(2\pi)^2\hbar^3} \epsilon^{1/2}.
\end{equation}

In the former case, namely $k<\frac{2\pi}{L}$, instead, the volume to be considered is obtained by subtracting the intersection volume of the red sphere with the two white spheres from the red sphere. 

A simple calculation yields:
\begin{equation}
    Vol_k = \frac{4\pi k^3}{3} - V_{inter} = \frac{Lk^4}{2} 
    \label{case2}
\end{equation}
where
\begin{equation}
V_{inter}=\frac{4\pi k^3}{3} - \frac{Lk^4}{2}
\end{equation}
is the intersection volume between the red sphere and the two white spheres of forbidden states in Fig.\ref{fig:Fig3}(a).
Equation~\eqref{case2} expresses the total volume of accessible states in $k$-space when $k<\frac{2\pi}{L}$, and thus can be used in order to find the total number of accessible states by using Eq.\eqref{eq:N}.
The corresponding DOS takes the following non-trivial form, 
\begin{equation}
\begin{split}
    N(k'< k) &= \frac{V}{(2\pi)^3}  \frac{L k^4}{ 2 }, \\
    N(\epsilon'< \epsilon) &= \frac{V}{(2\pi)^3}  \frac{L (2m\epsilon)^2}{2 \hbar^4}, \\
    g(\epsilon) &= \frac{d}{d\epsilon} N(\epsilon'<\epsilon) = \frac{VL m^2}{2\pi^3 \hbar ^4 } \epsilon.
    \label{eq:den2}
\end{split}
\end{equation}

Considering the two regimes depicted in Fig.\ref{fig:Fig3}, the overall DOS can be finally expressed as:
\begin{equation}
    g(\epsilon)=\begin{cases}   \frac{VL m^2}{2\pi^3 \hbar ^4 } \epsilon, & \mbox{if }  \epsilon < \frac{2\pi^2 \hbar^2}{mL^2} \\ \frac{V (2m)^{3/2}}{(2\pi)^2(\hbar)^3} \epsilon^{1/2}, & \mbox{if } \epsilon > \frac{2\pi^2 \hbar^2}{mL^2}.
\end{cases}
\label{eq:g}
\end{equation}
In reality, it is possible that there is a smooth crossover between the two regimes, which may also depend on the detailed system-specific boundary conditions of the sample and which cannot be determined within our analytical approach.

A similar derivation, \emph{mutatis mutandis}, for the case of phonons in confined solids has been presented in recent work in \cite{omega3}, where it has been successfully validated against both Molecular Dynamics (MD) simulations and experimental data based on inelastic neutron scattering.

As a final note, in the limit $L\rightarrow 0$, it should be possible to rigorously demonstrate that the red region of available states in Fig. \ref{fig:Fig3}(a) should theoretically shrink to a circular surface $\pi k^{2}$ in the $k_{x}-k_{y}$ plane. Thus, when $L=0$, this $\sim k^{2}$ contribution which survives in the r.h.s. of Eq. \eqref{case2} in the asymptotic limit, would then lead one to recover the exactly 2D, well-known result $g(\epsilon)\sim const$, and $T_{c}=0$. As currently we do not have a mathematically rigorous proof for this limit, we should defer its detailed discussion to future work.

\section{BEC critical temperature under confinement}
\subsection{Solution scheme}
In order to find the critical temperature $T_c$ in the standard theory of Bose Einstein condensation (BEC), it is necessary to evaluate the integral:
\begin{equation}
    \int_0^\infty \frac{g(\epsilon)}{ze^{\beta\epsilon}-1}d\epsilon
    \label{eq:BE}
\end{equation}
where $z = e^{\beta \mu}$ is the fugacity, with $\mu$ the chemical potential.
In an unconfined system, this integral is easily solved by using the Gamma and Zeta functions. In a confined system, the DOS will be different, and will be determined by Eq. \eqref{eq:g} (multiplied by the appropriate spin factor):
\begin{equation}
    g(\epsilon)= g_{spin} \cdot \begin{cases}   \frac{VL m^2}{2\pi^3 \hbar ^4 } \epsilon, & \mbox{if }  \epsilon < \epsilon^* \\ \frac{V (2m)^{3/2}}{(2\pi)^2(\hbar)^3} \epsilon^{1/2} & \mbox{if } \epsilon > \epsilon^* 
\end{cases}
\label{modified_DOS}
\end{equation}

The factor $g_{spin}$ accounts for the total spin degeneracy given by $(2S+1)$, and hence it depends on the specific bosonic (system) which is taken into account. Without loss of generality, we consider $g_s=1$, therefore the following discussion is appropriate for spinless bosons. In order to consider bosons with spin, it is sufficient to multiply the final result by the appropriate spin degeneracy factor. 

As the next step to account for confinement, the evaluation of \eqref{eq:BE} has to be separated in the two intervals determined by the crossover energy $\epsilon^*$. 
Even without solving the integral, it is easy to see that the expected result consists in an increase of the critical temperature $T_c$ as confinement is turned on. This is a consequence of the fact that, if some low-energy states are prohibited, the macroscopic occupation of the ground state will necessarily begin at higher temperatures, since otherwise particles would occupy high energy states, for which the Boltzmann factor would be very small at low temperatures. \cite{BEC_GrossmannHolthaus}. 

\subsection{Evaluation of the Bose integral with the modified DOS}
The integral over the BE distribution, using the modified DOS of Eq.\eqref{modified_DOS}, can be expressed as:
\begin{equation}
\begin{split}
      \int_0^\infty \frac{g(\epsilon)}{e^{\beta\epsilon}-1}d\epsilon&=\int_0^{\epsilon^*} \frac{VL m^2}{2\pi^3 \hbar ^4 }  \frac{\epsilon}{e^{\beta\epsilon}-1}d\epsilon +\\
      &+\int_{\epsilon^*}^{\infty} \frac{V (2m)^{3/2}}{(2\pi)^2\hbar^3} \epsilon^{1/2} \frac{\sqrt{\epsilon}}{e^{\beta\epsilon}-1}d\epsilon.
\end{split}
\end{equation}

By changing variable to $x=\beta\epsilon$, one obtains:
\begin{equation}
\begin{split}
      \int_0^\infty \frac{g(\epsilon)}{e^{\beta\epsilon}-1}d\epsilon&=\int_0^{\beta\epsilon^*}\frac{VL m^2}{2\pi^3 \hbar ^4 \beta^2} \frac{x}{e^{x}-1}dx +\\
      &+\int_{\beta\epsilon^*}^{\infty} \frac{V (2m)^{3/2}}{(2\pi)^2\hbar^3 \beta^{3/2}}  \frac{\sqrt{x}}{e^{x}-1}dx
\end{split}
      \label{eq:integral}
\end{equation}
To lighten the notation, we define the following quantities: $A\equiv\frac{L m^2}{2\pi^3 \hbar ^4 \beta^2}$ and $B\equiv\frac{(2m)^{3/2}}{(2\pi)^2\hbar^3 \beta^{3/2}}$. 
This integral cannot be solved exactly, even with the aid of special functions, because the integral of $\frac{\sqrt{x}}{e^x-1}$ does not have analytical solutions (except for the definite integral from $0$ to $\infty$, which gives the well known textbook result for the BEC critical temperature). The second integral, with the factor $\frac{x}{e^x-1}$, on the other hand, can be integrated, and this will become useful in the following.  Its indefinite integral has the following solution:
\begin{equation}
    \int \frac{x}{e^x-1}dx = x\log(1-e^{-x})-Li_2(e^{-x}), \label{eq:indefintegral}
\end{equation}
where the function $Li_2(x)= \sum_{k=1}^{\infty} \frac{x^k}{k^2}$ is the base-2 polylogarithmic function, related to the Riemann's Zeta function via: 
\begin{equation}
    Li_n(1)=\zeta(n) \longrightarrow Li_2(1)=\zeta(2)=\frac{\pi^2}{6}.
    \label{eq:polilog-zeta}
\end{equation}

Although the above integral cannot be solved exactly, approximate solutions can be found in the different limits.  
A possible choice for the small parameter is given by
\begin{equation}
\beta\epsilon^*= \frac{1}{k_BT}\frac{2\pi^2\hbar^2}{mL^2}.
\label{key_parameter}
\end{equation}
If this parameter is $\ll 1$, namely if $L$ is sufficiently large, then the first integral in Eq.\eqref{eq:integral} can be neglected, and calculations give the bulk value for critical temperature $T_{c,\infty}$, i.e. the standard BEC textbook result \cite{huang}. 
If the parameter$\beta\epsilon^*$ is slightly increased by decreasing $L$, approximate calculations can be performed by Taylor expanding the arguments of the two integrals, and this will be done in the next Section\ref{section:approx_solutions}. 

The opposite limit consists in the limit of large $\beta\epsilon^*$ (hence the small parameter is $1/\beta\epsilon^*$), which is obtained for very small values of $L$: in this limit, the second integral on the right hand side of \eqref{eq:integral} can be neglected, and a new formula for critical temperature will be obtained, as will be shown and explained in Section \ref{section:low_l}.

In the following sections we shall therefore obtain results in the limits discussed above.

\subsection{General form of the confinement-induced corrections}
\label{section:approx_solutions}
 In order to Taylor-expand the integrand functions, it is useful to write:
\begin{equation}
     \int_{\beta\epsilon^*}^{\infty}  \frac{\sqrt{x}}{e^{x}-1}dx = \int_{0}^{\infty} \frac{\sqrt{x}}{e^{x}-1}dx 
     - \int_{0}^{\beta\epsilon^*} \frac{\sqrt{x}}{e^{x}-1}dx
     \label{eq:split}
\end{equation}
where the common prefactor $BV$ has been omitted.

The integral in \eqref{eq:integral} can then be rewritten as:
\begin{equation}
\begin{split}
      &\int_0^\infty \frac{g(\epsilon)}{e^{\beta\epsilon}-1}d\epsilon=\int_{0}^{\infty} B\,V \frac{\sqrt{x}}{e^{x}-1}dx +\\ &+\left(\int_0^{\beta\epsilon^*} A\,V \frac{x}{e^{x}-1}dx - \int_{0}^{\beta\epsilon^*} B\,V \frac{\sqrt{x}}{e^{x}-1}dx \right)
\end{split}
    \label{eq:correction}
\end{equation}
The first integral in Eq.\eqref{eq:correction} is the usual integral that has to be solved in order to find the critical temperature of a non-confined condensate, therefore it will give the standard result, namely $\frac{V}{\lambda^{3}}\zeta(\frac{3}{2})$, where $\zeta$ is the Riemann zeta function and $\lambda$ is the thermal (de Broglie) wavelength. The terms in bracket provide, instead, the total correction to the integral due to confinement, therefore they generate a correction in the value of $T_c$ compared to the unconfined result $T_{c,\infty}$. Moreover, the two correction integrals have now the same extremes of integration, therefore they can be treated using the same approximations: this will prove particularly useful in calculations of the moderate-$L$ (small $\beta \epsilon^{*}$) regime of the theory.

\section{Approximate analytical solutions}
\subsection{First-order calculations in the low-to-moderate confinement limit}
\label{section:fo}
The first limit that can be investigated is the limit for which the parameter $\beta \epsilon^*$ is small. This is the limit of moderate or sufficiently large $L$: a precise evaluation of what can be taken as ``large'' will be performed in Section \ref{section:parameters}.

If $\beta \epsilon^*$ is small, the denominators in the corrective integrals in Eq.\eqref{eq:correction} can be Taylor expanded to the desired order, and the integrals become easily solvable (this can be done because the integrals are taken from $0$ to $\beta\epsilon^*$, hence the usefulness of the manipulation Eq.\eqref{eq:split} is now clear). 
By Taylor expanding the exponentials in the integrals to the first order, $e^x \simeq 1+x$, one gets:
\begin{gather}
    \label{eq:folinear}
    \int_0^{\beta\epsilon^*}AV\frac{x}{e^x-1}dx \simeq  \int_0^{\beta\epsilon^*}AVdx = AV\beta\epsilon^* = \frac{Vm}{\pi\hbar^2L\beta}\nonumber\\
    \label{eq:fosqrt}
    \int_0^{\beta\epsilon^*}B\,V\frac{\sqrt{x}}{e^x-1}dx \simeq  \int_0^{\beta\epsilon^*}\frac{BV}{x^{1/2}}dx = \frac{2\sqrt{\epsilon^*}}{\beta} = \frac{2Vm}{\pi\hbar^2L\beta}.
\end{gather}

The total correction is given by the difference of the two integrals, namely $-\frac{Vm}{\pi\hbar^2L\beta}$, as clear from Eq.\eqref{eq:correction}. As is standard \cite{pahria1977,huang}, the critical temperature $T_c$ for BEC can be found by considering the total number of particles as:
\begin{equation}
    N = \left \langle n_0 \right \rangle + \int_0^\infty \frac{g(\epsilon)}{e^{\beta\epsilon}-1}d\epsilon,
\end{equation}
where the integral can be expressed as Eq.\eqref{eq:correction}. Solving the integral gives:
\begin{equation}
     N = \left \langle n_0 \right \rangle +\frac{V}{\lambda^3}Li_{3/2}(1)-\frac{Vm}{\pi\hbar^2L\beta}.
\end{equation}
The last two terms are, respectively, the solution to the standard integral (first integral in Eq. \eqref{eq:correction}), which gives the standard solution for bulk condensates, $\frac{V}{\lambda^{3}}\zeta(\frac{3}{2})$, and the solution of the correction integrals just obtained. Calculations can now be carried on similarly to what is done in the standard, bulk or unconfined, case:
\begin{gather}
    n = \frac{\left \langle n_0 \right \rangle}{V} + \frac{Li_{3/2}(1)}{\lambda^3}-\frac{m}{\pi\hbar^2L\beta}, \\ 
    \frac{\lambda^3\left \langle n_0 \right \rangle}{V}= \lambda^3n +\frac{m\lambda^3}{\pi\hbar^2L\beta}-Li_{3/2}(1)
\end{gather}
where in the second line we multiplied through by the cube of the de Broglie thermal wavelength $\lambda = \hbar \sqrt{\frac{2\pi}{mk_{B}T}}$.
The critical wavelength $\lambda_{c}$ (and thus the critical temperature $T_c$) can be obtained by setting the right hand side of the above equation to zero:
\begin{equation}
    \lambda_c^3 \, n +\frac{m\lambda_c^3}{\pi\hbar^2L\beta_c}-Li_{3/2}(1) = 0.
    \label{eq:firstorder}
\end{equation}
This formula can be rearranged to show the behaviour of the critical temperature $T_c$ or of the critical wavelength $\lambda_c$ as functions of the thickness of the film $L$.

Since $\lambda_c=\hbar\sqrt{\frac{2\pi}{mK_BT_c}}$, Eq. \eqref{eq:firstorder} can easily be rewritten as:
\begin{equation}
    \lambda_c^3 \, n+\frac{2\lambda_c}{L}=Li_ {3/2}(1).
    \label{eq:folambda}
\end{equation}
The second term, $\frac{2\lambda_c}{L}$, is then a corrective term which depends on the thickness of the confined film. It is evident that the effect of this correction leads to a decrease in the value of $\lambda_c$ with respect to the bulk value $\lambda_{c,\infty}$, since a new, positive term is added to the standard $n\cdot\lambda_c^3$ term. This equation can be solved exactly by using Cardano's formula, which leads to the following explicit solution:
\begin{equation}
\begin{split}
    \lambda_c=&\sqrt[3]{\frac{Li_{3/2}(1)}{2n}+\sqrt{\frac{Li_{3/2}^2(1)}{4n^2}+\frac{8}{27n^3L^3}}} \\ &+\sqrt[3]{\frac{Li_{3/2}(1)}{2n}-\sqrt{\frac{Li_{3/2}^2(1)}{4n^2}+\frac{8}{27n^3L^3}}}~.
    \end{split}
    \label{eq:cardano}
\end{equation}

\begin{figure}[htbp]
    \centering
    \includegraphics[width=.45\textwidth]{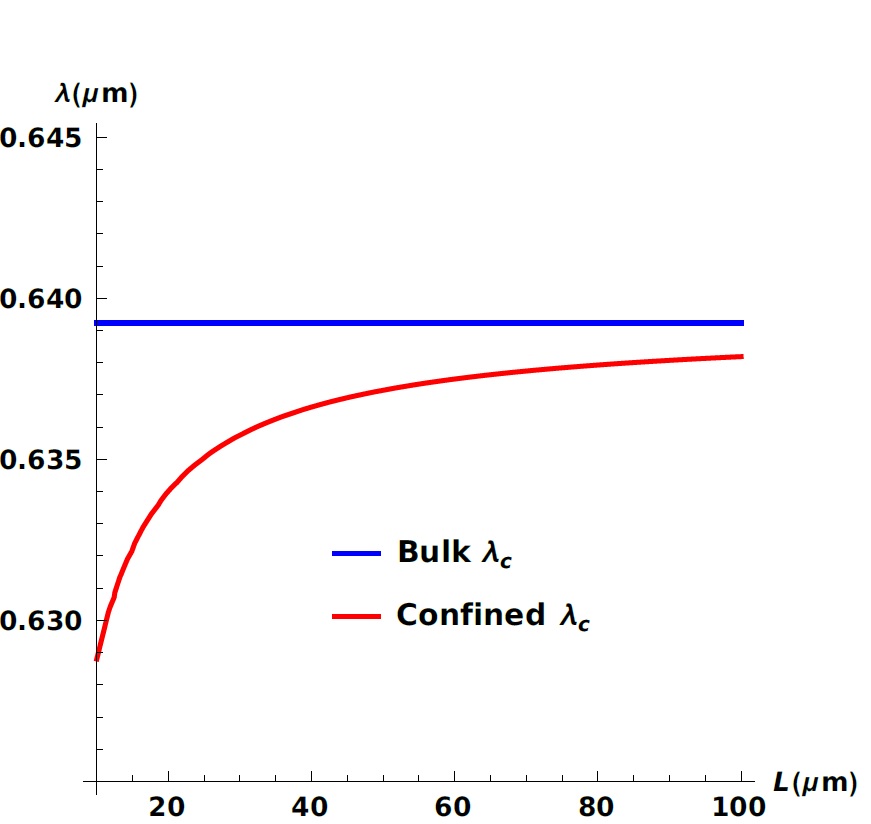}
    \caption{First-order calculations of the critical thermal wavelength $\lambda_{c}$, valid for moderate confinement ($L$ values not too small such that $\beta \epsilon^{*} \ll 1$) based on Eq.\eqref{eq:cardano} show that $\lambda_c$ decreases as $L$ decreases. All the physical parameters used in the calculations are taken from the case of $^{87}$Rb, see e.g. Ref.\cite{first_BEC}. The blue line indicates the critical thermal wavelength for the $^{87}$Rb system of Ref.\cite{first_BEC}.}
    \label{fig:folambda}
\end{figure}

As shown in Fig. \ref{fig:folambda}, the critical thermal wavelength $\lambda_c$ calculated with the above formula decreases with respect to the bulk value as the film  thickness $L$ is decreased. 

A more compact (compared to Eq.\eqref{eq:cardano}) approximate expression for $\lambda_c$ in this first order limit can be obtained by evaluating Eq. \eqref{eq:folambda}. This is done, again, by considering $\lambda_c \approx \lambda_{c,\infty} + \delta\lambda_{c}$, where $\delta\lambda_{c}$ is considered a small parameter, such that second and higher order terms in $\delta\lambda_{c}$ can be neglected. This is justified by the fact that $\lambda_c$ does not change too drastically with respect to the bulk value in the range of application of the approximations used in this section. Substituting in Eq. \eqref{eq:folambda}, one obtains:

\begin{equation}
\begin{split}
    &\lambda_c\left(n\lambda_c^2+\frac{2}{L}\right)= Li_ {3/2}(1) \\
    &(\lambda_{c,\infty}+\delta\lambda_{c})\left(n(\lambda_{c,\infty}+\delta\lambda_{c})^2+\frac{2}{L}\right) =  Li_ {3/2}(1).
\end{split}
\end{equation}
Upon developing the square and using $n\lambda_{c,\infty}^{3}= Li_{3/2}(1)$, we then obtain:
\begin{equation}
\delta\lambda_{c} = -\frac{2\lambda_{c,\infty}}{L}\, \frac{1}{\left(3n\lambda_{c,\infty}^{2}+\frac{2}{L}\right)}
\end{equation}

Therefore, an approximated form of Eq. \eqref{eq:cardano} is given by:
\begin{equation}
\lambda_c = \lambda_{c,\infty} + \delta\lambda_{c} = \lambda_{c,\infty}\left(1-\frac{2}{2+3nL\lambda_{c,\infty}^{2}}\right)
\label{eq:lambdaapproxapprox}
\end{equation}

This equation provides a more compact but still highly accurate approximation of the exact solution of Eq. \eqref{eq:folambda}, which can be implemented more easily in calculations and further analytical developments of the theory. This expression shows that the main consequence of confinement is a decrease in the value of $\lambda_c$ with respect to $\lambda_{c,\infty}$, and that the magnitude of the decrease is $\propto L^{-1}$.

\begin{figure}[htbp]
    \centering
    \includegraphics[width=.45\textwidth]{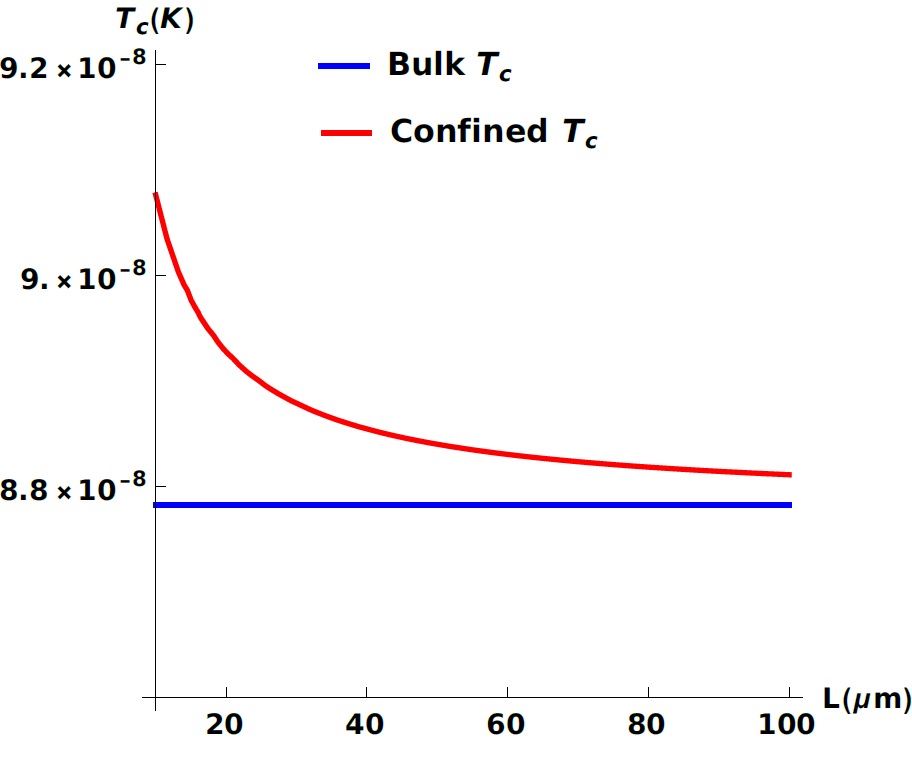}
    \caption{The critical temperature $T_c$, computed based on Eq.\eqref{eq:tcfo}, increases significantly as the thickness $L$ of the film is decreased from large values ($L\rightarrow \infty$)  to  $L\simeq 10^{-5}$m. The curves are plotted using values of the physical parameters tabulated for $^{87}$Rb cfr. Ref.\cite{first_BEC}. }
    \label{fig:fotc}
\end{figure}

Since temperature is related to $\lambda$ via $T=\hbar^2\frac{2\pi}{mk_B\lambda^2}$, the critical temperature $T_c$ will do just the opposite, i.e. it will increase as the value of $L$ is decreased. The closed-form expression which describes the behaviour of $T_c$ as a function of $L$ can be found by inverting Eq. \eqref{eq:folambda}, and substituting $\lambda_c = \hbar \sqrt{\frac{2\pi}{mk_bT_c}}$:

\begin{gather}
    n\,\hbar^3\left(\frac{2\pi}{mk_{B}T_c}\right)^{3/2} + \frac{2\hbar}{L} \sqrt{\frac{2\pi}{mk_{B}T_c}} =Li_ {3/2}(1),\nonumber
\end{gather}

\begin{equation}
    n\, \hbar^3\left(\frac{2\pi}{mk_B} \right)^{3/2} + \frac{2\hbar}{L} \sqrt{\frac{2\pi}{mk_{B}}}\,T_c - Li_{3/2}(1)T_c^{3/2} = 0.    \label{eq:tcfo}
\end{equation}
The exact formula for the $T_c$ as a function of $L$ can be found by inversion of Eq.\eqref{eq:cardano}. This leads to a cumbersome and not particularly illuminating solution; it is therefore more useful to solve the implicit Eq. \eqref{eq:tcfo} graphically or numerically, obtaining the result shown in Fig.\ref{fig:fotc}. In the range where this approximation is valid, the $T_c$ shows a slight increase with respect to the bulk value upon increasing the confinement, i.e. upon decreasing $L$.
It is interesting to notice the fact that the effect of confinement becomes relevant at much higher values of $L$, compared to the case of critical temperature for superconductivity in BCS-type superconductors \cite{goodcomp}. This can be explained by a simple consideration: the effect of confinement will have a relevant effect when the thickness $L$ of the film becomes comparable with the thermal wavelength of bosons, which at such low temperatures is of the order of a few $\mu$m. Hence, the increase in $T_c$ will be significant for much higher $L$ compared to the situation of electrons in a superconductor, where the thermal wavelength is around 1 nm.

Qualitative information about the behaviour of $T_c$ as a function of $L$ in this limit can be extracted by inversion of Eq. \eqref{eq:lambdaapproxapprox}. By considering $T_c \propto \lambda_c^{-2}$, one has:
\begin{equation}
\begin{split}
    T_c &= \frac{T_{c,\infty}}{\left(1-\frac{2}{2+3nL\lambda_{c,\infty}^{2}}\right)^2}\\
    &\approx {T_{c,\infty}} \left(1+\frac{4}{2+3nL\lambda_{c,\infty}^{2}}\right).
    \end{split}
\end{equation}
From this expression it is clear that also the $T_c$ goes as $\propto L^{-1}$ (at leading order), and it increases with respect to the bulk value upon increasing the confinement, thus confirming analytically the results shown in Fig. \ref{fig:fotc}.

\subsection{Second-order calculations in the moderate confinement limit}
\label{section:so}
    The results obtained in the previous section can be improved by expanding the exponential to the second order. This leads to higher precision in numerical solutions although, as it will be clear shortly, the correction is almost negligible with respect to the first order correction in the region in which the Taylor expansion is justified. It is worth noting that, at this order, the equations obtained for $\lambda_c$ are still solvable by using Cardano's formula Eq.\eqref{eq:cardano}, while higher orders do not provide analytically solvable equations (since the equations would involve fifth- or higher-order polynomials). 
    
To second-order in $\beta \epsilon^{*}$ we thus have for the first piece of the Bose integral:
    \begin{gather}
    \label{eq:solinear}
        \int_0^{\beta\epsilon^*}\frac{x}{e^x-1}dx \simeq  \int_0^{\beta\epsilon^*}\frac{x}{x+\frac{x^2}{2}}dx \\
        =  \int_0^{\beta\epsilon^*}\frac{1}{1+\frac{x}{2}}dx \simeq  \int_0^{\beta\epsilon^*}(1-\frac{x}{2})dx \\
      =\beta\epsilon^*-\frac{(\beta\epsilon^*)^2}{4}
    \end{gather}
and, similarly, for the second piece:
\begin{gather}
\label{eq:sosqrt}
    \int_0^{\beta\epsilon^*}\frac{\sqrt{x}}{e^x-1}dx \simeq  \int_0^{\beta\epsilon^*}\frac{\sqrt{x}}{x+\frac{x^2}{2}}dx  \\
    =\int_0^{\beta\epsilon^*}\frac{1}{\sqrt{x}}\frac{1}{1+\frac{x}{2}}dx \simeq  \int_0^{\beta\epsilon^*}\frac{1}{\sqrt{x}}(1-\frac{x}{2})dx \\
   = 2\sqrt{\beta \epsilon^*}-\frac{(\beta \epsilon^*)^{3/2}}{3}.
\end{gather}

The total correction is then given by:
\begin{equation}
    A\, V \int_0^{\beta\epsilon^*}\frac{x}{e^x-1}dx-B \, V\int_0^{\beta\epsilon^*}\frac{\sqrt{x}}{e^x-1}dx= -\frac{Vm}{\pi\hbar^2L\beta}+\frac{V\pi}{6L^3}   
\end{equation}

This new equation gives a further correction to Eq. \eqref{eq:firstorder}, which can be evaluated as:
\begin{gather}
    N = \langle n_0 \rangle + \frac{V}{\lambda^3}Li_{3/2}(1)-\frac{Vm}{\pi\hbar^2L\beta}+ \frac{V\pi}{6L^3} \\
    \frac{\lambda^3 \langle n_0 \rangle}{V} = \lambda^3 n + \frac{m\lambda^3}{\pi\hbar^2L\beta}-Li_{3/2}(1)-\frac{\pi\lambda^3}{6L^3} \\
    (n-\frac{\pi}{6L^3})\lambda_c^3+2\frac{\lambda_c}{L}=Li_{3/2}(1).
    \label{eq:sotc}
\end{gather}
Equation \eqref{eq:sotc} provides a further correction to the critical wavelength and hence to the $T_c$. In the limit now considered, this correction term is very much smaller than the first correction term found in the previous Section, because of the dependence on $L^{-3}$: as clear from the term in parenthesis, this correction would become relevant for values $L^{-3}$ comparable to the density of bosons, but this does not happen in the case of $L\approx 10^{-6}$m.

\begin{figure}
    \centering
    \includegraphics[width=.49\textwidth]{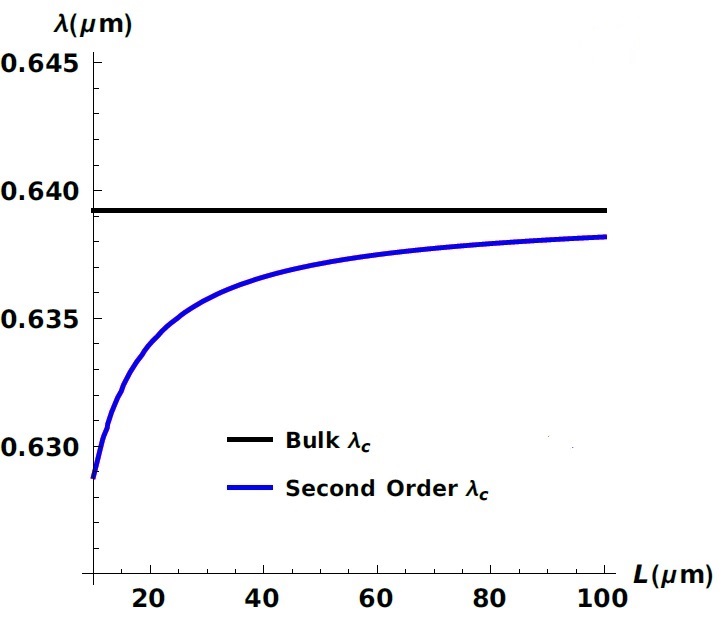}
    \caption{Bulk value, first and second order calculations for critical wavelength. The first and second order calculations are completely indistinguishable on the scale of the plot. 
     Therefore, from a practical point of view, the first order equation \eqref{eq:folambda} provides an excellent approximation to the exact solution of the integrals, since further expansion of the exponential will give even smaller corrections (this is valid for sufficiently high values of $L$: as it will be discussed in Section \ref{section:parameters}, if $L \approx 5 \cdot 10^{-6}$ the error in the first order approximation becomes increasingly significant, so the second order calculation becomes necessary, but that is not particularly useful since for $L \leq 5 \cdot 10^{-6}$ the approximation valid in the low-$L$ is more useful and sufficiently accurate. }
    \label{fig:second_order}
\end{figure}

Therefore, as clear from Fig.\ref{fig:second_order}, the second order correction causes a negligible increase in the value of $\lambda_c$ (hence a very slight decrease in the value of the critical temperature) compared to what is predicted by first order calculations. This provides a corroboration of the validity of the approximation made, since higher order corrections will provide even smaller deviations from the curves shown in Fig. \ref{fig:second_order}. The main reason to consider second-order corrections would be to maintain high precision in the approximation when decreasing the value of $L$ to only a few nanometers; clearly, performing calculations at increasingly higher orders can eventually extend the validity of the Taylor expansion to any values of $L$, but this would have no actual usefulness since at low values of $L$ one can use the simpler form of the approximation valid in the low-$L$ regime that we will develop in the next Section.

Similar calculations as in the previous section can be done, in order to find again a simple and compact solution of Eq. \eqref{eq:sotc} without going through solving a cubic equation. In fact, calculations are exactly the same, the only difference being that $n$ is replaced by $(n-\frac{\pi}{6L^3})$,
\begin{equation}
    \begin{split}
        &\lambda_c\left[(n-\frac{\pi}{6L^3})\lambda_c^2+\frac{2}{L}\right]= Li_ {3/2}(1) \\
    &(\lambda_{c,\infty}+\delta\lambda_{c})\left[(n-\frac{\pi}{6L^3})(\lambda_{c,\infty}+\delta\lambda_{c})^2+\frac{2}{L}\right] =  Li_ {3/2}(1),
    \end{split}
\end{equation}
where in the second line we  used $\delta \lambda_{c} = \lambda_{c} - \lambda_{c,\infty}$. Upon developing the square and using $n\lambda_{c,\infty}^{3}= Li_{3/2}(1)$, we obtain:
\begin{equation}
\delta\lambda_{c} = \frac{\frac{\pi\lambda_{c,\infty}^{3}}{6L^3}-\frac{2\lambda_{c,\infty}}{L}} {{\left[3(n-\frac{\pi}{6L^3})\lambda_{c,\infty}^{2}+\frac{2}{L}\right]}}.
\end{equation}
This equation is no longer as simple as the one found in the calculations made in the previous section, but provides a slightly better approximation if more precise results are needed.

\subsection{The low-$L$, strong confinement limit}
\label{section:low_l}
If $L$ is taken small enough, i.e. on the nanometer scale, then the parameter $\beta\epsilon^*$ will not be small enough to justify the calculations made in the previous sections. Therefore, a new approach is needed to investigate the behaviour of bosons confined in nanometer-scale thin films. Considering the integrals in Eq.\eqref{eq:integral}, it can be noted that the integrand function goes to zero exponentially fast as $x \rightarrow \infty$. Hence, the second integral $\int_{\beta\epsilon^*}^\infty \frac{\sqrt{x}}{e^x-1}$ quickly becomes negligible as $L \rightarrow 0$, and the first integral is well approximated by an integral going up to infinity. Hence, one has:
\begin{equation}
    \int_0^\infty \frac{g(\epsilon)}{e^x-1}d\epsilon \simeq \int_0^{\beta\epsilon^*} AV \frac{x}{e^x-1}dx \simeq \int_0^\infty AV \frac{x}{e^x-1}dx.
\end{equation}

This integral can easily be solved using the formula in Eq.\eqref{eq:indefintegral}, and using the fact that $Li_2(1)=\zeta(2)=\frac{\pi^2}{6}$. This gives
\begin{equation}
\begin{split}
      A\,V\int_0^\infty\frac{x}{e^x-1}dx &=  AV \left[x\log(1-e^{-x})-Li_2(e^{-x})\right]_0^\infty\\
      &= A\,V\, \frac{\pi^2}{6}.
\end{split}
\end{equation}

Contrarily to what was done in the opposite limit, this is not a correction to the standard calculation, but it is the full Bose integral. The total number of particles can be expressed as:
\begin{gather}
    N= \langle n_0 \rangle +A\,V\, \frac{\pi^2}{6}  = \langle n_0 \rangle +\frac{\pi^2}{6}\frac{VLm^2}{2\pi^3\hbar^4\beta^2},\\
    \frac{\langle n_0 \rangle}{V} = n - \frac{\pi^2}{6}\frac{Lm^2}{2\pi^3\hbar^4\beta^2}.
\end{gather}

As seen before, the critical temperature can be found by setting the right hand side to zero. By solving for $T_c$ one obtains
\begin{gather}
    T_c^2 = \frac{12 \pi \hbar^4 n}{L m^2 k_B^2}.
\end{gather}

\begin{figure}
    \centering
    \includegraphics[width=.49\textwidth]{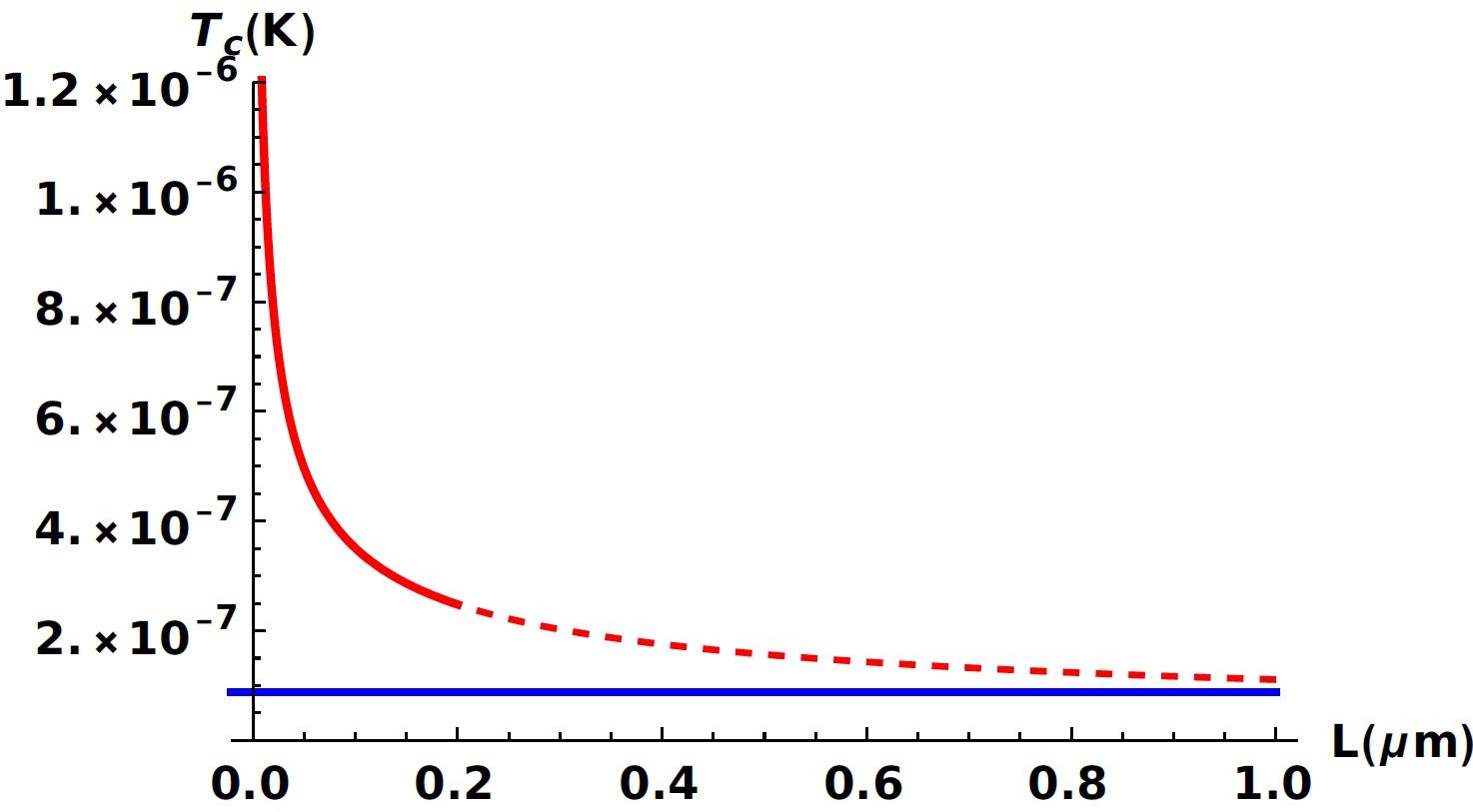}
    \caption{The critical temperature $T_c$ shows a substantial increase as $L$ becomes smaller than $10^{-7}$m, reaching around ten times the bulk value for $L \approx 10^{-7}$m. The dashed red line shows the values of $L$ for which the assumption of small $\beta\epsilon^*$ is not so well justified, and so the approximations made may not be very precise. The blue line represents the bulk value $T_{c,\infty}$.}
    \label{fig:LowL}
\end{figure}

This leads to two compact expressions for the critical values of temperature and wavelength:
\begin{gather}
    \lambda_c=\left(\frac{\pi L}{3n}\right)^{1/4}\nonumber\\
    T_c = \frac{2\hbar^2}{m k_B}\sqrt{\frac{3n\pi}{L}}.
    \label{key}
\end{gather}
This is a key result of this paper, i.e. the prediction of a square-root divergence $T_{c} \sim L^{-1/2}$ for strongly confined thin films. The corresponding predictions, using parameters of the $^{87}$Rb cold atomic gas \cite{first_BEC}, are shown in Fig. \ref{fig:LowL}.

\subsection{Implications for high-temperature BEC}
Again, the critical temperature $T_c$ is predicted to increase as $L$ is decreased. 
According to this formula, Eq. \eqref{key}, high-temperature BEC could be possible in certain systems, in particular light atoms or quasi-particle excitations \cite{magnonsPRL} due to the presence of mass $m$ in the denominator, and for moderate to large densities due to the presence of the number density $n$ in the square root in the numerator.

For example, for $^87$Rb, a 0.4 nanometer thick gaseous film at a density of $n=10^{15}$cm$^{-3}$, according to the above formula, would exhibit a critical temperature $T_c =5.4 \cdot 10^{-5}$K. This is about two orders of magnitude higher than the typical $T_c$ of cold atomic gases which is in the order of hundreds of nanoKelvins or a couple of microKelvins at most. 
Extrapolating (with obvious caveats as our theory neglects interactions) to superfluid $^{4}$He  (mass density $125$kg m$^{-3}$), this would give $T_c \approx 6.5$K, for a 1 nm thick film and $T_c \approx 10.22$K for a 0.4 nm thick film.
Since $^{4}$He liquefies at about $4.2$K, this implies that the onset temperature for superfluidity under confinement can reach values that are close to the liquefaction temperature $\approx 4$K, thus higher than the typical regime $T_{c}<2.7$K where superfluidity of bosonic helium is observed.

Finally, it should be noted that the above Eq.\eqref{key} is obtained in the limit of $\beta \epsilon^{*} \gg 1$. Hence one could argue that this formula may lose accuracy at high temperatures. However, it should be noted that $\beta\epsilon^{*}$ goes as $\propto L^{-2}$ (see above Eq. \eqref{key_parameter}), which means that $\beta \epsilon^{*}$ grows faster with decreasing $L$ than it decays with increasing $T$, which should make this approximation quite robust also upon significantly increasing the considered temperatures as the film thickness decreases.

\subsection{Comparison with BCS superconductivity}
Interestingly, this low-$L$ behavior with $T_c$ monotonically increasing as $L$ is reduced towards the 2D limit, is just opposite to what happens in BCS superconductors, where one has a peak or maximum in $T_c$ vs $L$ and, for vanishing $L$, the critical temperature is observed to drop to zero \cite{goodcomp,valentinis}. This different behaviour is due to the fact that bosons do not occupy a Fermi sphere, not having to satisfy Pauli's principle, and the $T_c$ is not controlled just by the DOS evaluated at the energy of highest occupied level (Fermi), as is the case of BCS superconductors.
As shown in another paper focused on BCS theory, if we interpret Fig.\ref{fig:Fig3}(a) as the momentum space of fermions, it is clear that the DOS on the surface decreases upon further decreasing $L$ down to below a critical value of $L$ at which we have a transition from the situation of Fig.\ref{fig:Fig3}(b) to that of Fig.\ref{fig:Fig3}(a). Hence, one indeed observes the $T_c$ for superconductivity to increase with $L$
 up to a peak or maximum in superconducting thin films where BCS theory applies \cite{goodcomp,goodcomp2,valentinis}.
One should also notice, however, that, in real-world confined superconductors, the confinement does not affect only the electron density of states, but can also change the phonon properties, as discussed in \cite{Tamura1993,Bose2010,Croitoru2016} and references therein.

\section{Heat capacity}
Together with providing a correction to the BEC critical temperature, confinement will also cause significant modifications in the thermodynamic properties of the condensate, such as the condensate fraction (fraction of bosons in the ground state at a given temperature $T$) and the heat capacity of the system. While in the low-to-moderate confinement regime of micron-thick films these corrections may be difficult to observe, for sub-micron films where the approximations made in Section \ref{section:low_l} apply, the  effect will be much more significant. Evaluating the heat capacity of the condensate, and possibly of other thermodynamic properties in the presence of confinement is very important, since it offers a further possibility for experimental validation of the theory. For details of actual experimental setups, see Refs. \cite{Ensher1996,Ku2012,Damm2016,Schmitt2021}.

\subsection{Moderate-$L$ regime}
For bulk condensates the heat capacity goes like $ C \propto T^{\frac{3}{2}}$. In order to find the heat capacity for the confined systems, it is first necessary to determine the internal energy, which, in general for Bose systems, can be written as:
\begin{equation}
    U = \int_0^\infty \frac{\epsilon}{e^{\beta(\epsilon-\mu)}-1} g(\epsilon) d\epsilon.
\end{equation}
Focusing on the condensate, i.e. considering only the $T<T_c$ situation, $\mu$ can be set to zero.

To  calculate the heat capacity in the moderate-$L$ (micron-size) regime we need to apply the same approximations considered in Section \ref{section:fo} to study the $T_c$ in that regime. Since it was seen that the effect of the second-order corrections is negligible for a wide range of values of $L$, we will only consider first-order approximations in the present section. Starting with the general expression for the internal energy:
\begin{equation}
    U =\frac{VLm^2}{2\pi^3\hbar^4} \int_0^{\epsilon^*} \frac{\epsilon^2}{e^{\beta\epsilon}-1} d\epsilon + \frac{V(2m)^{3/2}}{4\pi^2\hbar^3} \int_{\epsilon^*}^\infty \frac{\epsilon^{3/2}}{e^{\beta\epsilon}-1} d\epsilon,
\end{equation}
this can be rewritten using the technique shown in Eq. \eqref{eq:split}:
\begin{equation}
\begin{split}
     U &=\frac{VLm^2}{2\pi^3\hbar^4} \int_0^{\epsilon^*} \frac{\epsilon^2}{e^{\beta\epsilon}-1} d\epsilon +\\ &+\frac{V(2m)^{3/2}}{4\pi^2\hbar^3} \left[\int_{0}^\infty \frac{\epsilon^{3/2}}{e^{\beta\epsilon}-1} d\epsilon  
     -\int_{0}^{\epsilon^*} \frac{\epsilon^{3/2}}{e^{\beta\epsilon}-1}d\epsilon \right]\\
     &= \frac{V(2m)^{3/2}}{4\pi^2\hbar^3\beta^{5/2}} \int_{0}^\infty \frac{x^{3/2}}{e^{x}-1} dx +\\ &+\frac{VLm^2}{2\pi^3\hbar^4\beta^3}\int_0^{\beta\epsilon^*} \frac{x^2}{e^{x}-1} dx  
     - \frac{V(2m)^{3/2}}{4\pi^2\hbar^3\beta^{5/2}}  \int_{0}^{\beta\epsilon^*} \frac{x^{3/2}}{e^{x}-1}dx.
\end{split}
\end{equation}
The first integral can easily be solved by considering that:
\begin{equation}
\int_0^\infty \frac{x^{3/2}}{e^x-1}dx = \frac{3}{4}\sqrt{\pi}\zeta(5/2).
\end{equation}
Using again the first-order approximation $e^x-1 \approx x$ in the other two integrals, this leads to the expression:
\begin{equation}
    U = k_1T^{5/2} + \frac{VLm^2}{2\pi^3\hbar^4\beta^3}\int_0^{\beta\epsilon^*} x dx + \frac{V(2m)^{3/2}}{4\pi^2\hbar^3\beta^{5/2}} \int_0^{\beta\epsilon^*} \sqrt{x} dx,
\end{equation}
where $k_1 = \frac{3}{4}\sqrt{\pi} \zeta(5/2) \frac{V(2m)^(\frac{3}{2})}{4\pi^2\hbar^3}k_B^{5/2}$ is a constant which does not depend on the confinement $L$, but only on the mass of the bosons and on the normalization volume; the first term in the equation is thus the bulk value of $U$. Solving the other two integrals gives:
\begin{equation}
    U =  k_1T^{5/2} + \frac{2Vk_B\pi}{3L^3}T =  k_1T^{5/2} + k_2T.
\end{equation}
Since $k_2$ depends on $L$, the second term provides the correction due to confinement to the internal energy of the condensate in the limit of moderate $L$.
The heat capacity is obtained as usual by differentiating the internal energy with respect to temperature:
\begin{equation}
    C = \frac{5}{2} k_1 T^{3/2} + k_2(L).
    \label{eq:c_highl}
\end{equation}
with $k_2 = \frac{2Vk_B\pi}{3L^3}$.

Therefore, confinement in the sub-millimeter to micron range causes a (in principle) measurable change in the heat capacity of the Bose gas, which scales as $L^{-3}$, which is another result of this paper.

Written like this, Eq. \eqref{eq:c_highl} might appear questionable, as it violates the third law of thermodynamics by not vanishing when $T\rightarrow 0$. However, the apparent contradiction is solved by considering the fact that the calculations just performed are valid if the parameter $\beta\epsilon^*$ is small, and this is impossible to occur for $T=0$. Therefore, the approximate result is consistent, and should be valid in the range in which the $\beta\epsilon^{*}\ll 1$ applies.

However, the correction given by Eq. \eqref{eq:c_highl} might be challenging to measure experimentally, since for $L\approx 10 ^{-5}$m the constant $k_2= \frac{2Vk_B\pi}{3L^3}$ is extremely small, providing corrections of only about $10^{-8}$J/K per unit volume. A larger change in $C$ due to confinement will be found in the low-$L$ limit, as will be discussed in the next section.   

\subsection{Low-$L$ limit}
In this case, using the same approximations of Section \ref{section:low_l}, we have, for the internal energy:
\begin{gather}
    U = \frac{VLm^2}{2\pi^3\hbar^4} \int_0^{\beta\epsilon^*} \frac{\epsilon^2}{e^{\beta\epsilon}-1}d\epsilon+ \frac{V (2m)^{\frac{3}{2}}}{4\pi^2\hbar^3}\int_{\beta\epsilon^*}^\infty \frac{\epsilon^{3/2}}{e^{\beta\epsilon}-1}d\epsilon \approx \\
    \approx \frac{VLm^2}{2\pi^3\hbar^4} \int_0^{\infty} \frac{\epsilon^2}{e^{\beta\epsilon}-1}d\epsilon = \frac{VLm^2}{\pi^3\hbar^4\beta^3}\zeta(3)
\end{gather}
and the result
\begin{equation}
    \int_0^\infty \frac{x^2}{e^x-1}dx = 2 \zeta(3)
\end{equation}
was used. Therefore, one has $U \propto T^3$, and consequently $C \propto T^2$. In particular, one has:
\begin{equation}
C = \zeta(3) \, \frac{3VLm^2}{\pi^3\hbar^4}k_B^3 T^2
\end{equation}
Contrarily to what was seen in the previous section, the effect in this limit is much more relevant.
First of all, the heat capacity exhibits an altogether different dependence on temperature compared to standard BEC theory, i.e. $C \sim T^{2}$ instead of $C \sim T^{3/2}$.
Furthermore, there is an $L$-dependent prefactor in front of the leading $T$-dependent term, hence the effect of confinement should definitely be experimentally measurable.

Interestingly, our result agrees with the prediction $C \sim T^{2}$ that has been obtained with completely different methods (i.e. a more elaborate variational approach \`a la Feynman) in the context of very thin layers (just few atomic layers) of helium on graphite \cite{Campbell1}. This is an interesting observation, which seems to imply that our $\sim T^{2}$ result for nanometer-scale films may hold continuously down to the 2D-like bilayer and monolayer systems where it merges with the law of Ref. \cite{Campbell1}. 

\section{Condensate fraction}
\subsection{Moderate-$L$ regime}
Since it was seen that the effect of the second-order correction in the expansion parameter $\beta \epsilon^{*}$ is negligible for a wide range of $L$ values, we will only consider first-order approximation in the present section. The fraction of particles which occupy the ground state at a given temperature in the presence of confinement can be found by considering:
\begin{gather}
    N(\epsilon>0) = \frac{V}{\lambda^3}\zeta(3/2)-\frac{Vm}{\pi\hbar^2L\beta} = \frac{V}{\lambda^3}\zeta(3/2) - \frac{2V}{L\lambda^2} \\ 
    =\frac{V}{\lambda^3}\left(\zeta(3/2)-\frac{2\lambda}{L}\right)
\end{gather}
Now we use the fact that, in the present moderate-$L$ limit, from Eq. \eqref{eq:folambda} one has $\zeta(3/2)=  n\lambda_c^3+\frac{2\lambda_c}{L}$, therefore the equation above can be rewritten as:
\begin{equation}
\begin{split}
     N(\epsilon>0) &= \frac{V}{\lambda^3}\left(n\lambda_c^3+\frac{2\lambda_c}{L}-\frac{2\lambda}{L}\right) \\
     &= Vn \frac{\lambda_c^3}{\lambda^3} +\frac{2V}{L\lambda^3}\left(\lambda_c-\lambda \right )
\end{split}
\end{equation}
Now, considering the definition of $\lambda$, it is clear that $\frac{\lambda_c^3}{\lambda} = \left(\frac{T}{T_c}\right)^{\frac{3}{2}}$. Using this substitution, together with $nV = N$, the result is:
\begin{equation}
    N(\epsilon>0) = N\left(\frac{T}{T_c}\right)^{\frac{3}{2}} + \frac{2V}{L\lambda^3}\hbar \sqrt{\frac{2\pi}{mk_B}}\left(\frac{1}{\sqrt{T_c}}-\frac{1}{\sqrt{T}}\right).
\end{equation}

The first term is the standard term as in the bulk BEC theory, whereas the second term is a corrective term induced by confinement: it is obviously negative, since the condensate fraction is evaluated for $T<T_c$. Therefore, confinement induces a reduction in the number of particles occupying excited states, and therefore an increase in the number of particle in the ground state, compared to the results obtained in unconfined condensates at the same values of $T$.

\subsection{Low-$L$ regime}
In a bulk condensate, the number of particles in the ground states is predicted to have a dependence on temperature of the form $\langle n_0 \rangle = N \left(1 - \left(\frac{T}{T_c}\right)^{\frac{3}{2}}\right)$. In the case of strong confinement, having a compact formulation of $T_c$ allows us to evaluate the confinement-induced correction to the condensate fraction, which could also be used to experimentally verify the theory. The number of bosons in the excited states is:
\begin{equation}
      N(\epsilon > 0) =  \int_0^\infty\frac{g(\epsilon)}{e^{\beta\epsilon}-1}d\epsilon
  \end{equation}
Using the same steps and approximations valid in the strong confinement, low-$L$ regime, we obtain
\begin{gather}
    N(\epsilon>0) = \frac{\pi^2VL m^2}{12\pi^3\hbar^4} k_B^2T^2= \\
    = V n T^2 \frac{12\pi\hbar^4 n}{L m^2 k_B^2} = N \frac{T^2}{T_c^2}.
\end{gather}

The number of particles in the ground state with $\epsilon=0$ is simply:
\begin{equation}
    \langle n_0 \rangle = N \left(1 - \left(\frac{T}{T_c}\right)^2\right).
\end{equation}

Since, for $T<T_c$ one has $\left(\frac{T}{T_c}\right)^2<\left(\frac{T}{T_c}\right)^{\frac{3}{2}}$, the fraction of particles in the ground state is greater for thin films than for bulk condensates.  This is qualitatively in agreement with numerical results on the cubic box geometry where, similarly, an enhancement of the condensate fraction as a consequence of confinement was numerically predicted for Dirichlet and antiperiodic boundary conditions in Ref. \cite{pahria1977}.

\section{The expansion parameter $\beta\epsilon^{*}$}
\label{section:parameters}
As shown above, the key parameter for the Taylor expansions in the various limits that one has to consider in order to obtain approximate solutions is 
\begin{equation}
    \beta\epsilon^* = \frac{1}{k_B T} \frac{2\pi^2\hbar^2}{mL^2}.\nonumber
    \label{eq:betaepsilon}
\end{equation}

The first approximation, which was considered in Sections \ref{section:fo} and \ref{section:so}, required this parameter to be small, and hence $L$ to be sufficiently large. To get an idea of the orders of magnitude where this requirement applies, it is necessary to estimate the prefactor in Eq. \eqref{eq:betaepsilon}. In order to do so, it is possible to consider the example of rubidium, which was the first Bose gas in which condensation was observed experimentally, and which has mass $m\simeq 87 \mbox{ a.m.u.} $. For the temperature value, we consider the condensation temperature of bulk rubidium, namely $T \approx 10^{-7}$K. Thus we have:
\begin{equation}
    \beta\epsilon^* \approx \frac{1.13\cdot 10^{-12}}{L^2}
    \label{eq:approx}.
\end{equation}

Simple numerical estimates show that the error in the integral using the second-order expansion is around or smaller than 0.1 \%, if $\beta\epsilon^*=0.1$.
Hence, inverting Eq.\eqref{eq:approx}, it is easy to obtain that this approximation is justified for $L> 3 \cdot10^{-6} m$, with a 0.1\% precision. Considering just the first-order expansion leads to the necessity of considering slightly higher values of $L$, since the error is of more than 2.5 \% for $L= 3 \cdot10^{-6} m$.  
It is worth noting that for different atoms, this value will be slightly different because of differences in the masses of the particles: lower masses will increase the prefactor, and thus increase the minimum value of $L$.

The strong confinement low-$L$ approximation, developed in Section \ref{section:low_l}, is the opposite approximation of large $\beta\epsilon^*$, i.e. in this case the small parameter is $1/\beta\epsilon^*$. In this case, the value of the parameter has to be large enough in order to justify neglecting the term 
\begin{equation}
    \int_{\beta\epsilon^*}^\infty B\frac{\sqrt{x}}{e^x-1}dx.
\end{equation}
Clearly, the argument of the integral is exponentially suppressed as $x\rightarrow \infty$. However, the prefactor takes a value $ B =  \frac{(2m)^{3/2}}{(2\pi)^2\hbar^3\beta^{3/2}}\approx 5\cdot 10^{18} $ (for rubidium). The approximation is valid if $$\int_0^{\beta\epsilon^*} A \frac{x}{e^x-1}dx \gg \int_{\beta\epsilon^*}^\infty B \frac{\sqrt{x}}{e^x-1}dx$$ and $$
\int_0^{\beta\epsilon^*} A \frac{x}{e^x-1}dx \approx \int_0^{\infty} A \frac{x}{e^x-1}dx.$$ 

As we verified, one has already great precision if $\beta\epsilon^* \approx 20$. This leads to a value of $L < 2\cdot 10^{-7}$m, so this approximation can be used to study confinement in films within a broad range of thickness from a fraction of micrometer down to nanometer systems.

\section{Comparison with similar systems}
Unfortunately, to date there are no experimental studies that explore the effect of confinement along a single spatial direction on the equilibrium properties of Bose-Einstein condensates, or the relationship between the condensation temperature and the thickness of the film, apart from studies focusing on the 2D limit of superfluids with just a couple atomic layers thickness where the physics is much more complicated \cite{Williams_onset}. However, various theoretical and especially numerical studies have been performed to describe finite size effects on BEC \cite{Ginzburg1967,BEC_GrossmannHolthaus,noronha,BEC_confinement}.

Of these contributions, only the paper by Ginzburg and Kirzhnits \cite{Ginzburg1967} focuses on the thin film geometry.
In order to describe a quasi-2D confined system they basically neglected all transverse contributions to the propagators, and obtained the following correction $T_{c} \propto T_{c,\infty}\frac{d n^{1/3}}{\ln (L^{2} d n)}$, where $d$ is the width of the system in the $x$ and $y$ directions. Clearly, also in this case $T_c$ increases with confinement although the formula becomes no longer valid when $L < (1/dn)^{1/2}$.
This model is more approximate than the framework presented here, since in our case we do not arbitrarily neglect all transverse modes, and we do retain all modes which are accessible in the presence of confinement.

As in the case of BCS superconductivity~\cite{ThompsonBlatt,valentinis}, the numerical solutions are generally found by imposing boundary conditions, such as Dirichlet ``hard-wall'' conditions, or periodic/antiperiodic conditions, that the wavefunction must satisfy at the borders. All studies predict an increase in critical temperature upon reducing the size of the confined system: as discussed above, this is a direct consequence of the fact that, as some low energy states are not accessible, the macroscopic occupation of the ground state begins at higher temperatures.

Recent results shown in Ref. \cite{BEC_confinement}, consider the change in the critical temperature of a condensate when confined in a cubic box of side $L$ using the three classic boundary conditions (Dirichlet, periodic/antiperiodic). The results show clearly that the specific choice of boundary condition is irrelevant in the prediction of the behaviour of the critical temperature, as the predicted behaviour is the same with all three types of boundary conditions. This is very different from the case of superconductors, for which the choice of different boundary conditions produces radically different behaviours of the superconducting parameters. Although the discussion in Ref.\cite{BEC_confinement} is referred to a cubic box and not to a thin film, this provides an interesting comparison to the results obtained in this work, since in both situations the increase in the critical temperature has the same origin. The power of the present model lays in the fact that analytical solutions are provided, rather than only the numerical calculations. Furthermore, the thin film geometry has a higher potential for the achievement of high-temperature BEC as confining the system in just one spatial direction allows for keeping the system macroscopic as a whole. With the cubic box model this is not possible as increasing the confinement necessarily implies reducing the size of the system as a whole.


\section{Conclusion}
We presented a theory of Bose-Einstein condensation in thin films which is fully analytical and covers a broad range of confinement size, from sub-millimeter to microns and down to the nanometer scale.

Previous approaches were either based on making strong approximations such as neglecting transverse modes \cite{Ginzburg1967}, or based on numerical calculations with certain boundary conditions \cite{BEC_GrossmannHolthaus,BEC_confinement}.

In this paper we took a radically different approach and developed an exact analytical solution which takes into account the effect of confinement along one spatial dimension on the available volume of accessible states in momentum space. For free bosons the available volume in momentum space is just a $k$-sphere, but in the presence of confinement this is no longer true and two spheres of forbidden states develop inside the $k$-sphere and their size grows with decreasing the film thickness $L$, as illustrated in Fig.\ref{fig:Fig3}.
This geometric distortion of the available momentum space has profound consequences on the density of states (DOS), and we were able to analytically evaluate the corrected DOS for thin-film confinement.

Analytical solutions for the critical temperature $T_c$ were developed for two distinct regimes, a moderate confinement regime valid for sub-millimeter films down to microns, and a strong confinement regime valid from fractions of microns down to the nanometer scale. In both regimes the critical temperature $T_{c}$ is predicted to increase with decreasing the film thickness, $L$. In the sub-millimeter to microns regime there is an additional term that contributes to $T_c$ and decays with $L^{-1}$.
In the strong confinement regime the $T_c$ is predicted to diverge with increasing confinement as $T_c \sim L^{-1/2}$. This implies that in this regime it should be possible to enhance the $T_c$ observed in cold atomic gases by at least two orders of magnitude with respect to the current experimental techniques. An enhancement up to about its liquefaction temperature should be observable also for the superfluidity onset temperature in helium films within the nanomteter thickness range.

Analytical predictions are also made for the heat capacity of the condensate.
In particular, in the strong confinement regime it is found that $C \sim T^{2}$, which is strikingly different from the standard BEC result $C \sim T^{3/2}$ and may have important consequences: this stronger increase of the heat capacity coming from the condensate may imply a different $\lambda$-point behavior, which may have observable consequences not only for dilute gases but also for superfluids such as $^{4}$He where the thin film geometry is easier to implement. 

These results open up new directions for a fundamental understanding of BEC and superfluidity onset and the enhancement thereof, with new testable predictions. In particular, it could be interesting to extend the confinement description of BEC presented here to predictions of the superfluid critical temperature using existing approximate theories \cite{Apenko}.\\

\textit{Acknowledgements} A.Z. acknowledges financial support from US Army Research Laboratory and US Army Research Office through contract nr. W911NF-19-2-0055. 


\bibliography{Bib}

\end{document}